# Topological descriptors for the electron density of inorganic solids

Nathan J. Szymanski[1], Alexander Smith[1], Prodromos Daoutidis[1], and Christopher J. Bartel[1,*]

[1] University of Minnesota, Department of Chemical Engineering and Materials Science, Minneapolis, MN, USA 55455
[*] Correspondence to cbartel@umn.edu

## Abstract

Descriptors play an important role in data-driven materials design. While most descriptors of crystalline materials emphasize structure and composition, they often neglect the electron density – a complex yet fundamental quantity that governs material properties. Here, we introduce Betti curves as topological descriptors that compress electron densities into compact representations. Derived from persistent homology, Betti curves capture bonding characteristics by encoding components, cycles, and voids across varied electron density thresholds. Machine learning models trained on Betti curves outperform those trained on raw electron densities by an average of 33 percentage points in classifying structure prototypes, predicting thermodynamic stability, and distinguishing metals from non-metals. Shannon entropy calculations reveal that Betti curves retain comparable information content to electron density while requiring two orders of magnitude less data. By combining expressive power with compact representation, Betti curves highlight the potential of topological data analysis to advance materials design.

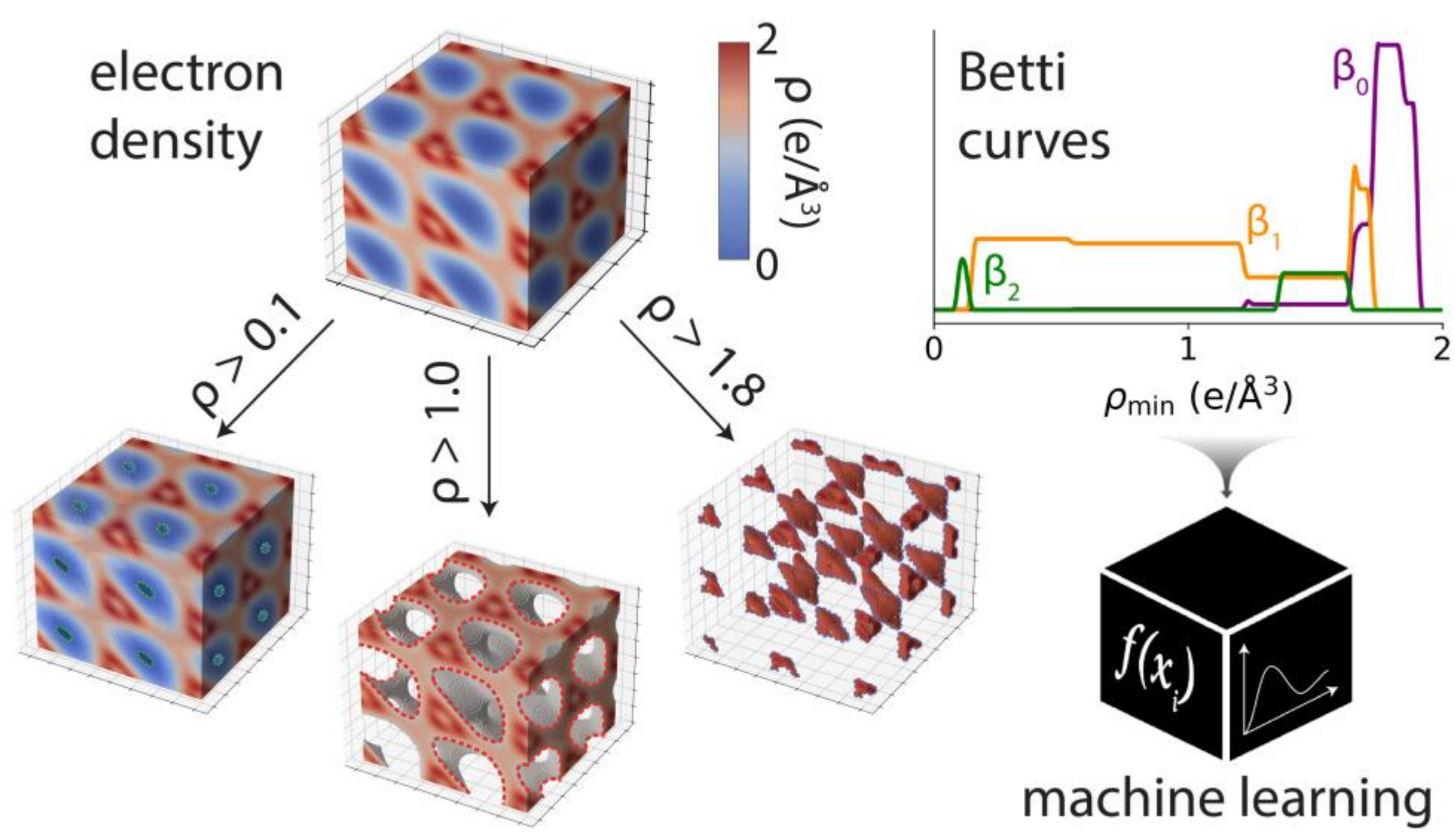

Machine learning (ML) is often enabled by compact numerical descriptors. In materials science, these descriptors encode structural and chemical information used as input to models that predict properties otherwise costly to compute with *ab initio* methods.[1] Popular examples include structural fingerprints,[2,3] which aggregate the properties of local coordination environments, smooth overlap of Gaussian-smeared atomic densities,[4] and feature embeddings learned from graph neural networks.[5–8] When paired with ML models trained on density functional theory (DFT) calculations, these descriptors can be used to predict material properties such as formation energy, band gap, and elastic moduli.[9,10]

Existing material descriptors tend to overlook the electron density, which describes the probability of finding an electron at a given position within the unit cell of a crystal. As described in the first Hohenberg-Kohn theorem,[11] the ground-state electron density uniquely determines all properties of a material. A variety of methods have therefore been developed to extract chemically meaningful insights from DFT-computed electron density. A popular example is Bader analysis,[12] which partitions electron density into basins to quantify charge transfer between atoms. This method and related schemes, such as Hirshfeld and Voronoi partitioning,[13] simplify the electron density into a small set of scalar quantities that can inform our understanding of oxidation states and describe the nature of chemical bonds.[14,15]

An alternative approach is to train ML models directly on the electron density without any prior partitioning or simplification. These models take as input a discretized version of the electron density, represented as a three-dimensional grid of density values (*i.e.*, voxels) at uniformly spaced coordinates. Applications range from property prediction to generative artificial intelligence (AI).[16–19] However, the low information content per voxel in electron density data poses challenges. To extract meaningful patterns, complex models such as convolutional neural networks (CNNs) are often employed, which require large datasets to learn sufficiently expressive representations.

We propose Betti curves as a compact, information-rich representation of the electron density tailored for ML models. These curves describe the density through its topological invariants: the number of connected components ($\beta_0$), cycles ($\beta_1$), and enclosed voids ($\beta_2$).[20,21] Each number is calculated iteratively over manifolds of the electron density ($\rho$), generated by applying a sequence of density thresholds (). Starting from $\rho_{\min} = 2\ e/\text{Å}^3$, a manifold is created by excluding regions with $\rho$ < $\rho_{\min}$ and the invariants are computed for the resulting manifold (Supplementary Note 1). This process is repeated with smaller filtration steps ($\Delta\rho_{\min} = 0.0125\ e/\text{Å}^3$) until $\rho_{\min} = 0\ e/\text{Å}^3$is reached.

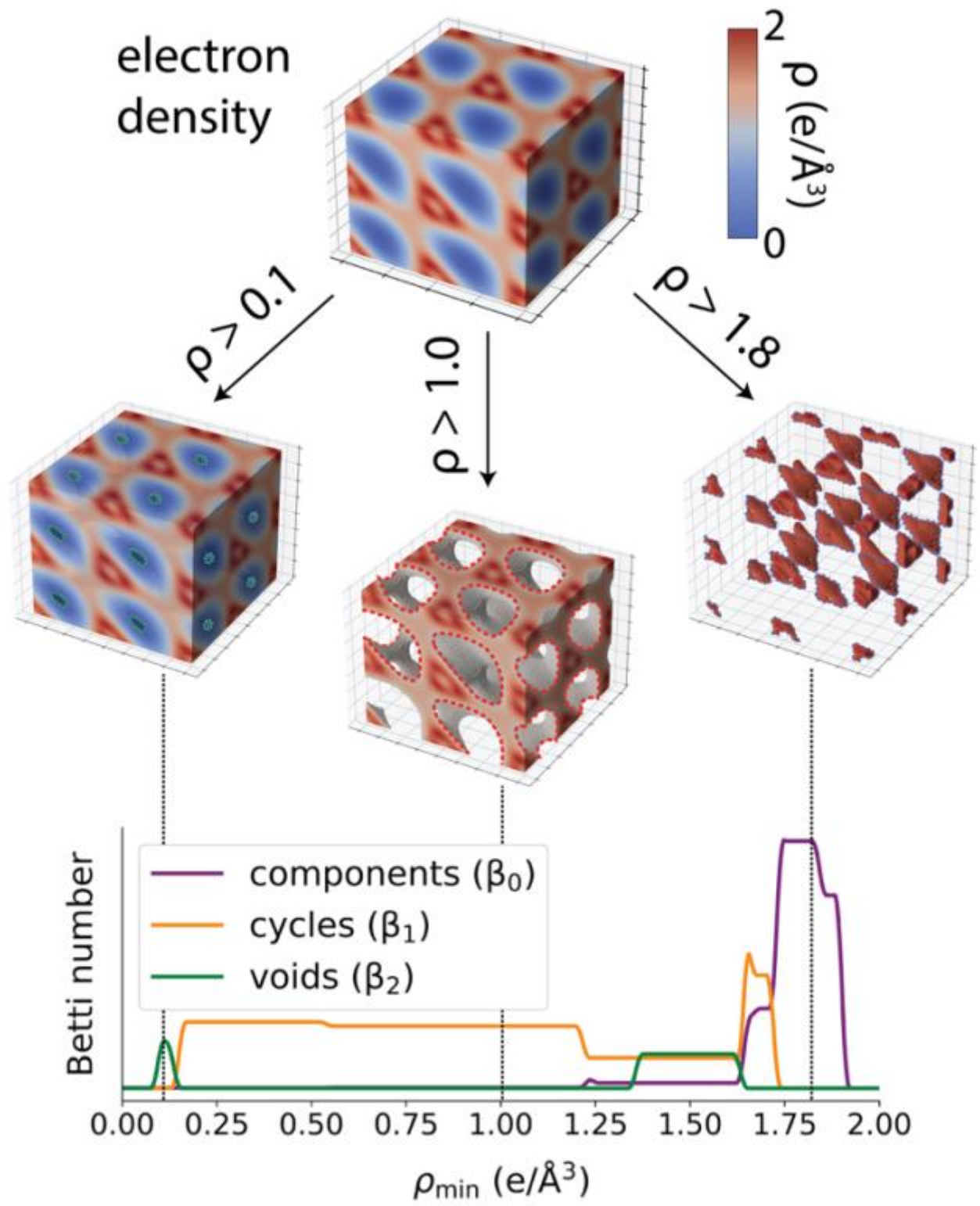

**Figure 1 | Generation of Betti curves from electron density.** (Top panel) Complete electron density of diamond obtained from DFT calculations. (Middle panels) Three filtrations of the electron density exceeding specific thresholds, ranging from $\rho > 0.1\ e/Å^3$ to $\rho > 1.8\ e/Å^3$. (Bottom panels) Betti curves generated from the electron density of diamond, with dashed vertical lines corresponding to sets of Betti numbers ($\beta_0$, $\beta_1$, $\beta_2$) computed at the varied thresholds ($\rho_{\text{min}}$).

The process described above produces a set of Betti curves – $\beta_0$, $\beta_1$, and $\beta_2$ – each consisting of 160 values that encode the evolution of topological features as the filtration threshold ($\rho_{\text{min}}$) decreases. Analysis of the electron density and Betti curves for diamond (**Figure 1**) reveals the below trends:

1. **Components ($\boldsymbol{\beta_0}$):** Large values of $\beta_0$ at high filtration thresholds correspond to localized regions of electron density near atomic centers. As $\rho_{\text{min}}$ decreases, these regions merge and $\beta_0$ decreases, reflecting increased connectivity in the electron density.
2. **Cycles ($\boldsymbol{\beta_1}$):** Cycles emerge as shared electron density forms "tunnels" between atoms. These features highlight increased bonding interactions and delocalized charge within the structure.
3. **Voids ($\boldsymbol{\beta_2}$):** At low filtration thresholds, $\beta_2$ represents enclosed voids in the electron density, corresponding to regions in the structure that are unoccupied by atoms or bonds.

To investigate these trends, we first examine how compositional changes in the rocksalt structure affect the Betti curves. This structure is one of the most common for binary compounds and spans a wide range of bonding characteristics.[22] As a case study, we focus on variations in covalency across three compositions: VO, VN, and VC. The integrated crystal orbital bond index (ICOBI), a measure of covalent bond strength (Supplementary Note 2),[23] shows that these compounds have increasing covalent character as the anion becomes less electronegative: 0.23 for VO, 0.27 for VN, and 0.29 for VC. Two-dimensional slices of their electron densities, along the (100) plane, are shown in the top panels of **Figure 2**. The black contours in each plot represent surfaces of equal density. The bottom panels contain the corresponding Betti curves of each rocksalt material, where vertical lines mark the electron density values associated with the contour lines in the top panels.

The bonding is predominantly ionic in VO, with electrons highly localized around individual ions (V and O). This localization is reflected in the Betti curves by many isolated components ($\beta_0$) over a wide range of electron density thresholds ($\rho_{\mathrm{min}}$). Minimal sharing of electrons between V and O leads to a narrow peak in the number of cycles ($\beta_1$) at approximately $0.1-0.3\ e/\text{Å}^3$, followed by a small peak in the number of voids ($\beta_2$) near zero electron density. In contrast, the bonds in VN are moderately covalent. While charge remains localized at high densities near the V and N atoms, increased electron sharing leads to a broader $\beta_1$ plateau that occupies $\rho_{\mathrm{min}}$ ranging from $0.2-0.5\ e/\text{Å}^3$. The covalent bond character in VN also introduces directionality in the electron distribution, creating additional voids (higher $\beta_2$) compared to VO. This trend continues with VC, where stronger covalent bonding leads to more substantial electron sharing between the V and C atoms. The result is an even wider $\beta_1$ plateau, ranging from $0.2-0.6\ e/\text{Å}^3$. Additional voids also emerge at high $\rho_{\mathrm{min}}$, as evidenced by the appearance of a new plateau in $\beta_2$ at values of $1.4-1.7\ e/\text{Å}^3$. These voids arise from the delocalized nature of charge around the central C atom, which spreads out and is shared with neighboring V cations. The resulting distribution creates a void within the C atom that is absent in VO and VN.

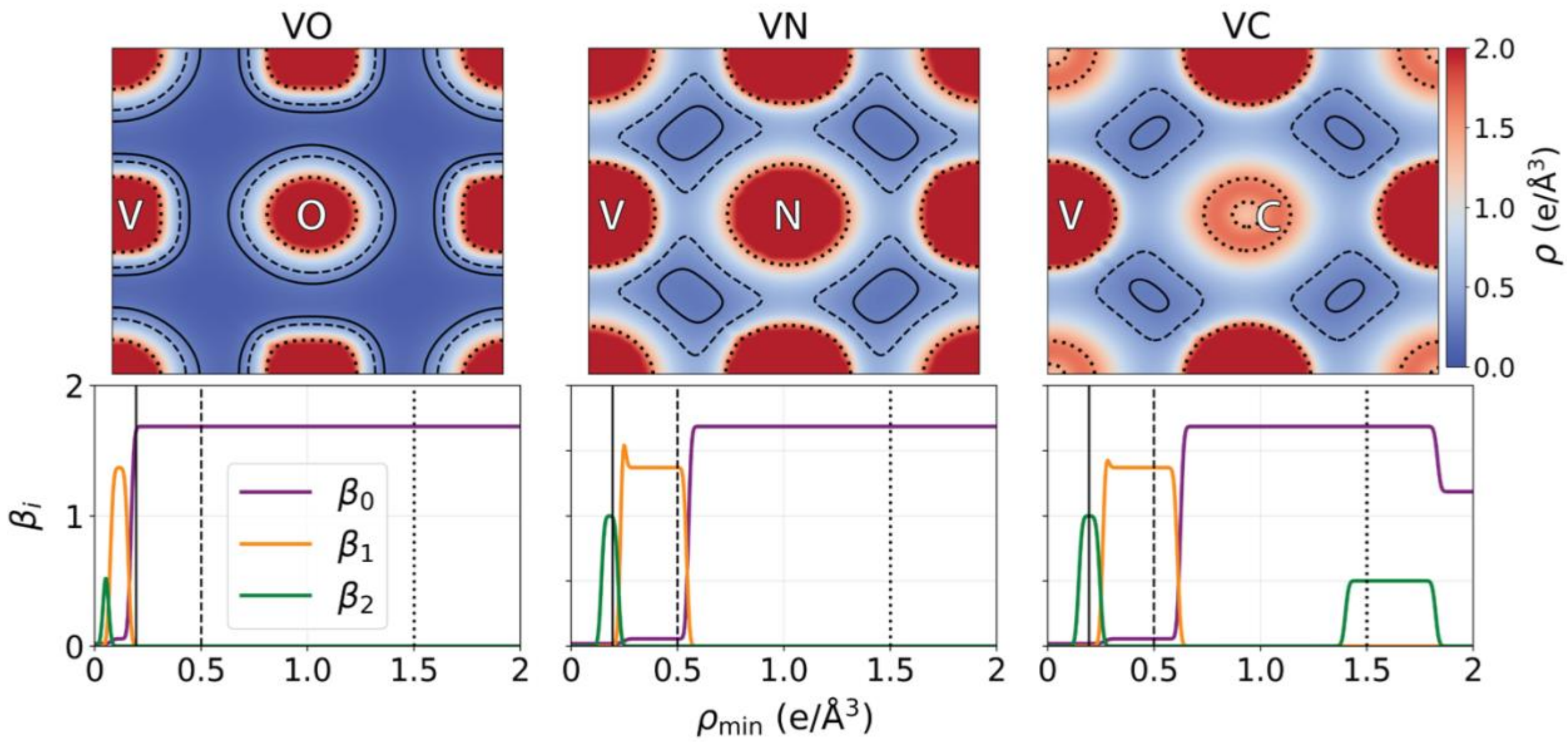

**Figure 2 | Electron densities and Betti curves of rocksalt materials.** (Top panels) Two-dimensional slices of the DFT-calculated electron density for VO, VN, and VC (left to right) are visualized along the (100) plane. Each material is modeled in a rocksalt structure. Solid, dashed, and dotted lines represent contours of equal density. (Bottom panels) Betti curves generated from these electron densities. Vertical lines connect the contours in the top panels to specific electron density thresholds ($\rho_{\mathrm{min}}$) in the Betti curves.

The most distinguishing feature in the Betti curves that separates ionic from covalent bonding is the width of non-zero values in $\beta_1$ (illustrated in **Figure 3a**). This width reflects the number of cycles in various manifolds of the electron density and indirectly captures the range of $\rho_{\mathrm{min}}$ over which charge is shared between distinct ions. Materials with greater electron sharing (*i.e.*, more covalent bonds) tend to exhibit broader $\beta_1$ widths, while those with minimal electron sharing (more ionic bonds) display narrower widths. To better illustrate this trend, **Figure 3b** presents a scatter plot where each point corresponds to a unique rocksalt material. The color of each point represents the width of $\beta_1$ in the Betti curves generated from that material. The x-axis shows the percentage ionic character of nearest-neighbor bonds in each rocksalt material, calculated from Bader charge analysis,[24,25] while the y-axis shows covalent bond strength indicated by ICOBI.[23] Details on the calculation of these quantities are provided in Supplementary Note 2. A comparison with percent charge transfer derived from Mulliken rather than Bader charge, is also provided in Supplementary Figure 1.

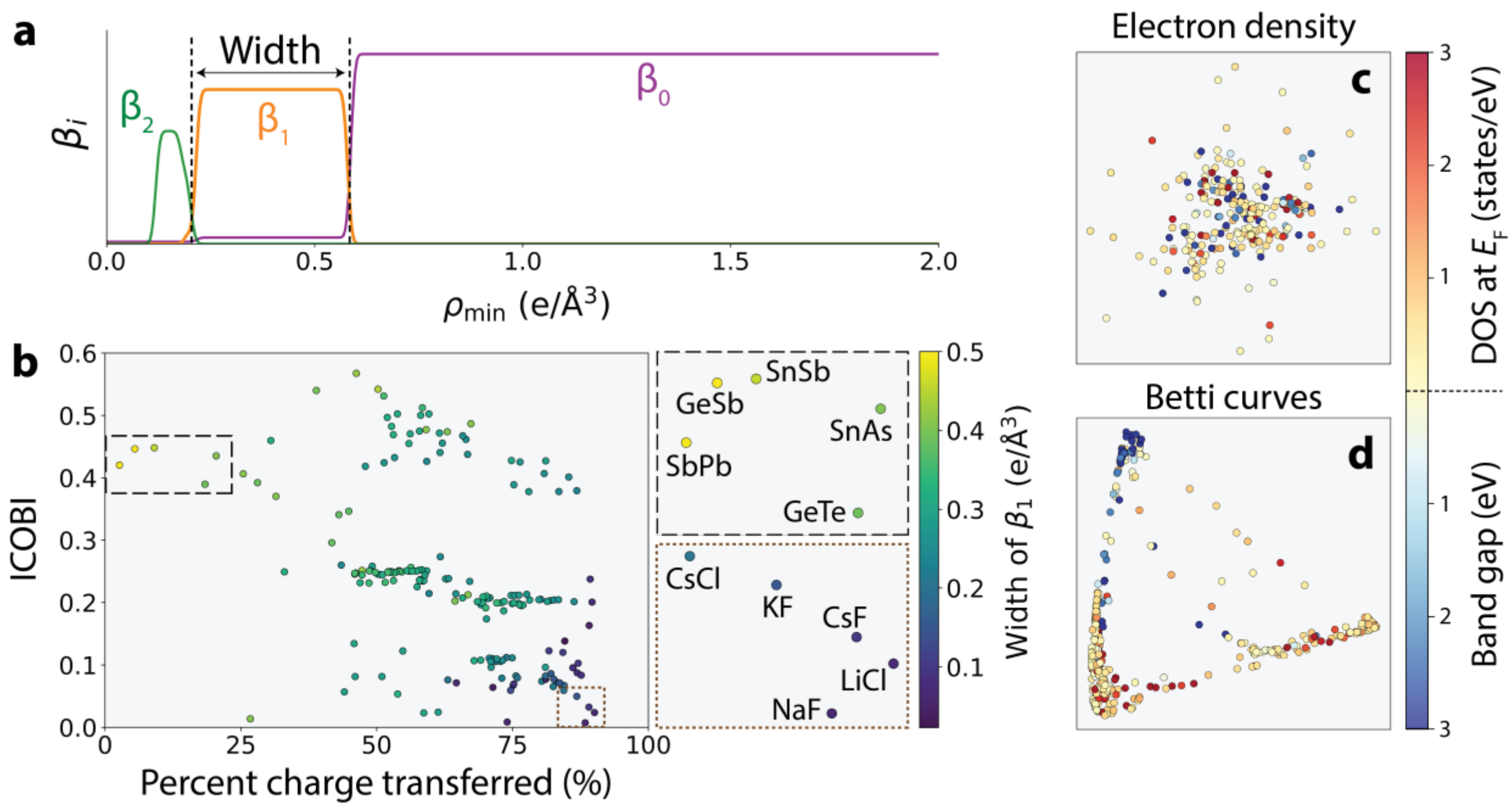

**Figure 3 | Relations between bond character and Betti curves in rocksalt materials. (a)** The width of non-zero values in $\beta_1$ (orange line) is determined from the Betti curves. **(b)** Scatter plot showing the relationship between the percent charge transferred (x-axis) and ICOBI (y-axis) for various rocksalt materials. Each point represents a distinct material, with its color indicating the width of $\beta_1$ in its Betti curves. (Bottom-right panels) Insets showing materials with highly covalent (top) and ionic (bottom) bond character. Right panels show the results of spectral embedding applied to **(c)** electron densities and **(d)** Betti curves of rocksalt materials. Points represent materials, color-coded by electronic band gap (for non-metals) or density of states (DOS) at the Fermi level (for metals).

The scatter plot reveals a clear relationship between the width of $\beta_1$ and the nature of bonding. Ionic materials, characterized by high charge transfer and low ICOBI values, cluster in the lower right of the plot. These materials exhibit narrow $\beta_1$ widths, indicative of limited electron sharing between cations and anions. For example, alkali halides such as LiCl and CsF exhibit narrow $\beta_1$ widths, as highlighted in the bottom-right inset of **Figure 3b**. In contrast, covalent materials with low charge transfer and high ICOBI values occupy the upper left of the plot. These materials display broad $\beta_1$ widths, reflecting substantial electron sharing between atoms. Group IV-V compounds (*e.g.*, GeSb and SnAs) are prime examples, as shown in the top-left inset of **Figure 3b**. The broad $\beta_1$ widths in these materials indicate a shared electron distribution between the group IV and V elements, consistent with their strong covalent bonding. These results demonstrate how the width of $\beta_1$ serves as a scalar descriptor for capturing the extent of charge sharing, offering a quantitative link between topological features of the electron density and the underlying bonding characteristics of materials.

While the width of non-zero values in $\beta_1$ is the most visually apparent feature linked to bonding, more subtle features in the Betti curves likely encode additional information. We used unsupervised learning to automatically extract these features and determine whether similar features could be learned directly from the electron density. **Figure 3c-d** illustrates the results of spectral embedding applied to electron densities (top panel) and Betti curves (bottom panel) of these same rocksalt materials. Each point corresponds to a unique material, color-coded by its band gap (for non-metals) or its density of states at the Fermi level (for metals).

When spectral embedding is applied directly to electron densities (**Figure 3c**), no discernible trends emerge. In contrast, applying spectral embedding to Betti curves reveals a well-defined triangular arrangement of points in the latent space (**Figure 3d**). Non-metals with large band gaps (blue dots) cluster at the apex of the triangle. Covalently bonded materials with small band gaps or low density of states at the Fermi level (yellow dots) are distributed along the edges or center of the triangle. Metallic materials with high density of states at the Fermi level (red dots) are concentrated near the bottom corners of the triangle. These results show that Betti curves capture features related to chemical bonding and electronic properties, enabling unsupervised learning to organize materials without requiring manual input.

We next explore how structure influences the appearance of Betti curves by examining three binary structure prototypes: CsCl, rocksalt, and zincblende. For each prototype, electron densities from all available entries in the Materials Project were used to generate Betti curves. **Figure 4** illustrates the distributions of Betti curves for materials in each prototype. The intensity of shading represents the concentration of curves, with darkest regions centered on the median values (shown by black lines) and gradually fading as curves deviate from the median. Above these distributions, a slice of the electron density is shown for representative compounds: CsCl for the CsCl structure type, NbC for rocksalt, and SiC for zincblende.

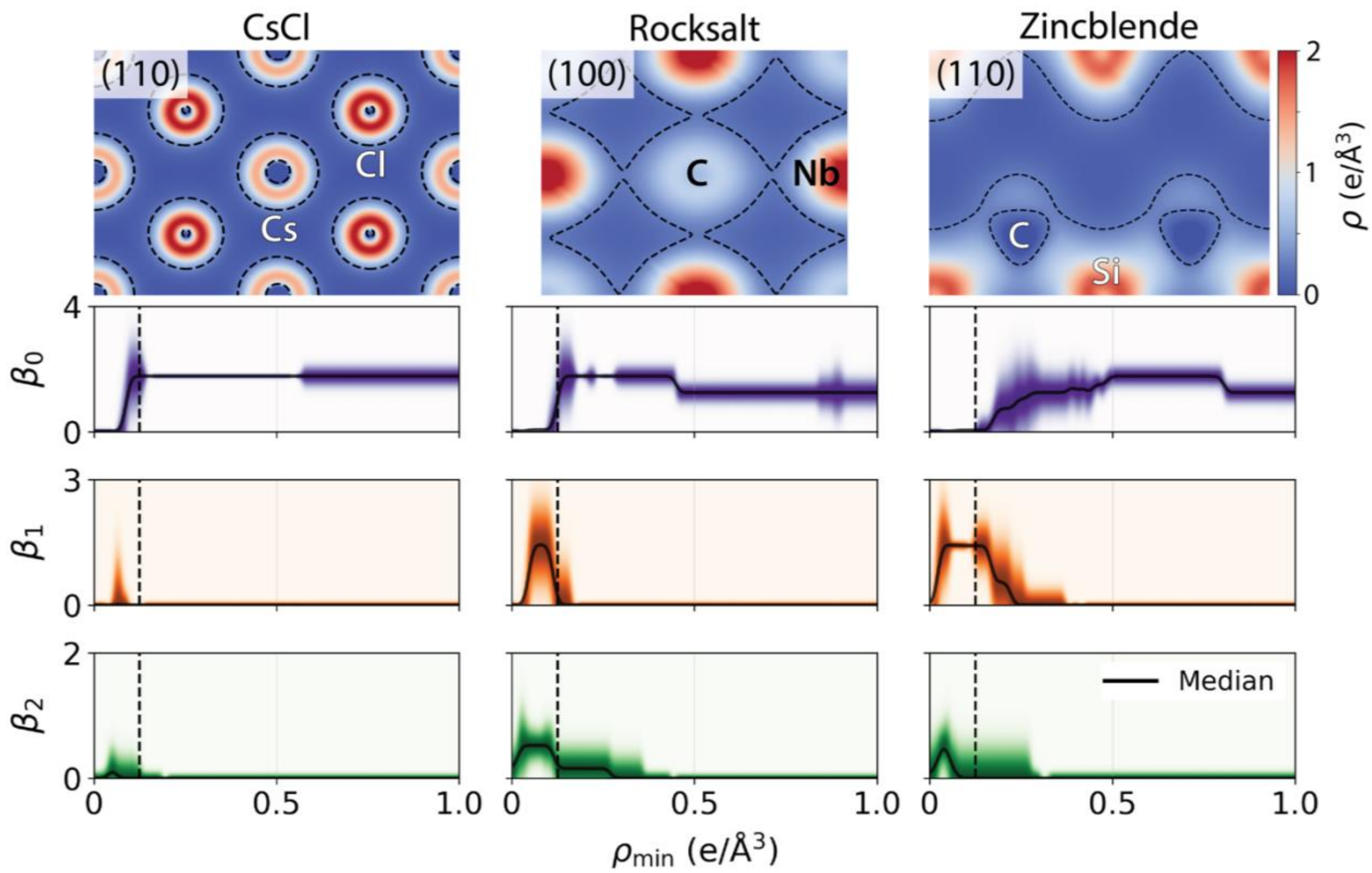

**Figure 4 | Electron densities and Betti curves of distinct structural prototypes.** Electron densities (top row) for CsCl, NbC, and SiC as representative materials in the CsCl, rocksalt, and zincblende prototypes (left to right). Distributions of Betti curves (bottom three rows) for materials in the CsCl, rocksalt, and zincblende prototypes. Color intensity shows the distribution of Betti curves for materials in each prototype, with the darkest regions at median values (black lines). Solid, dashed, and dotted lines represent contours of equal density. Dashed vertical lines connect the contours in the top panels to specific electron density thresholds in the Betti curves.

The Betti curves of CsCl-structured materials exhibit narrow $\beta_1$ peaks at low $\rho_{\min}$, reflecting minimal charge sharing typical of highly ionic compounds. These $\beta_1$ peaks separate high electron-density components ($\beta_0$) near ion cores from low-density voids ($\beta_2$) between ions, consistent with the strongly ionic nature of compounds such as CsBr and RbCl. In contrast, rocksalt materials show wider $\beta_1$ plateaus at slightly higher $\rho_{\min}$, indicative of the mixed ionic and covalent bonding characteristic of this prototype. Greater variations in the Betti curves (*i.e.*, wider distributions) also highlight the increased diversity of bonding character found in the rocksalt family.[22] Similar variations are observed in zincblende materials, as evidenced by the Betti curves shown in the rightmost panels of **Figure 4**. Wide $\beta_1$ plateaus often extend to high values of $\rho_{\min}$, signifying the presence of strong covalent bonds that would be anticipated for zincblende materials. The directional nature of this bonding also creates additional "pockets" of localized charge between atoms, which are reflected in the Betti curves as frequent and sharp transitions in $\beta_0$, corresponding to changes in the number of components as $\rho_{\min}$ varies.

The ability of Betti curves to distinguish structural prototypes aligns with long-standing efforts to classify binary compounds based on scalar descriptors.[1] For example, Pettifor introduced a chemical scale ($\chi$) to create a 2-D mapping of binary compounds based on their valence electron count, ionic core radii, and electronegativity.[26] More recent strategies have used features like Born effective charges and electronegativity differences to separate binary compounds by structure type.[27] Identification of optimal features to distinguish these structures types has also been automated through data-driven approaches such as compressed sensing.[28] Betti curves provide additional insight that complements existing descriptors, offering a compact topology-based representation of the electron density.

To evaluate whether Betti curves provide information that is directly useful to ML models, we trained three random forest classifiers, each with a distinct objective: identifying structure prototypes, distinguishing metals from non-metals, and predicting thermodynamic stability. For each task, we compare ML predictions using a variety of inputs: raw electron density, Betti curves generated from the electron density, structure fingerprints (SOAP,[4] Matminer,[2] and SkipAtom[3]), and pooled feature embeddings obtained from pre-trained graph neural networks (CHGNet[7] and MACE[29]). Details on these methods are provided in Supplementary Notes 3-5.

**Table 1 | Classification from a variety of descriptors.** Classification accuracies on a hold-out test set of 9,317 materials spanning 846 structural prototypes obtained from AFLOW.[30–33] Random forest classifiers were trained separately on electron densities (from the Materials Project),[34] Betti curves, SOAP features,[4] Matminer fingerprints,[2] SkipAtom vectors,[3] and pooled feature embeddings from graph neural networks (CHGNet and MACE).[7,29] Three classification tasks were evaluated: identifying structural prototypes, distinguishing metallic from non-metallic compounds, and classifying a material as thermodynamically stable or unstable.

| Descriptor | Structure | Metal/non-metal | Stability |
|---|---|---|---|
| Electron density | 21.0% | 60.8% | 69.8% |
| Betti curves | 84.7% | 85.3% | 79.6% |
| SOAP | 79.7% | 83.1% | 74.3% |
| Matminer | **87.2%** | 84.3% | 77.7% |
| SkipAtom | 59.6% | 84.6% | 72.5% |
| CHGNet | 79.1% | **90.1%** | **80.8%** |
| MACE | 60.3% | 89.6% | 78.4% |

We note that DFT calculations are required to generate electron densities and their Betti curve representations. If these calculations have already been performed on a material, then predicting its band gap or stability with ML is redundant. Likewise, if a structure is known from a DFT calculation, its prototype can be identified without using ML. As such, these experiments focus on assessing the predictive power of different representations rather than creating practically useful ML models. For each task described in the previous paragraph, we trained random forest classifiers using different representations as input, with results summarized in **Table 1**.

Random forest classifiers trained on Betti curves generally achieve high accuracies (> 79% across all tasks), showcasing the information encoded in these descriptors. Betti curve-based models consistently outperform those trained directly on raw electron densities, with an average accuracy improvement of about 33 percentage points across tasks. The improvement is most pronounced for structure classification, where training on Betti curves yields a four-fold increase in accuracy compared to models trained on raw electron densities.

Betti curves also perform comparably well to previously established descriptors. For structure classification, random forests trained on Betti curves underperform only those trained on representations from Matminer, which were specifically designed to serve as structure fingerprints by encoding information regarding local coordination environments.[2] In contrast, using Betti curves leads to better performance than structure fingerprints for the remaining tasks: distinguishing metals from non-metals and predicting thermodynamic stability. Pooled feature embeddings extracted from pre-trained graph neural networks (CHGNet and MACE) provide strong representations for property prediction but are less effective than Betti curves for structure classification. Notably, random forests trained on Betti curves achieve an average classification accuracy of 83.2% across all three tasks – comparable to the models trained on CHGNet (83.3%) and MACE (76.1%) embeddings.

The improved performance of classification models using Betti curves suggests that these descriptors encode information more efficiently than electron density, offering a compact format that is well suited for simple ML algorithms like random forests. To quantify the information content of each descriptor, we approximate its Shannon entropy using a kernel density estimate (KDE) approach outlined in previous work and described in Supplementary Note 6.[35,36] In the top panel of **Figure 5**, we plot the Shannon entropy of Betti curves and electron densities from all 93,178 materials used in the classification tasks described earlier. The entropy is computed as a function of each descriptor's sampling density, which measures how many scalar values it contains per sample. Betti curves and electron densities both show a monotonic increase in entropy with longer descriptor lengths, eventually plateauing near 10.5 nats. However, while the Betti curves reach this maximum entropy at relatively low sampling densities (~200 data points per sample), electron densities require much higher sampling densities (> 10,000 data points) to achieve the same plateau. This provides further evidence that Betti curves offer a more compact, information-rich representation of the electron density, compressing it more than 50× with little to no loss of information.

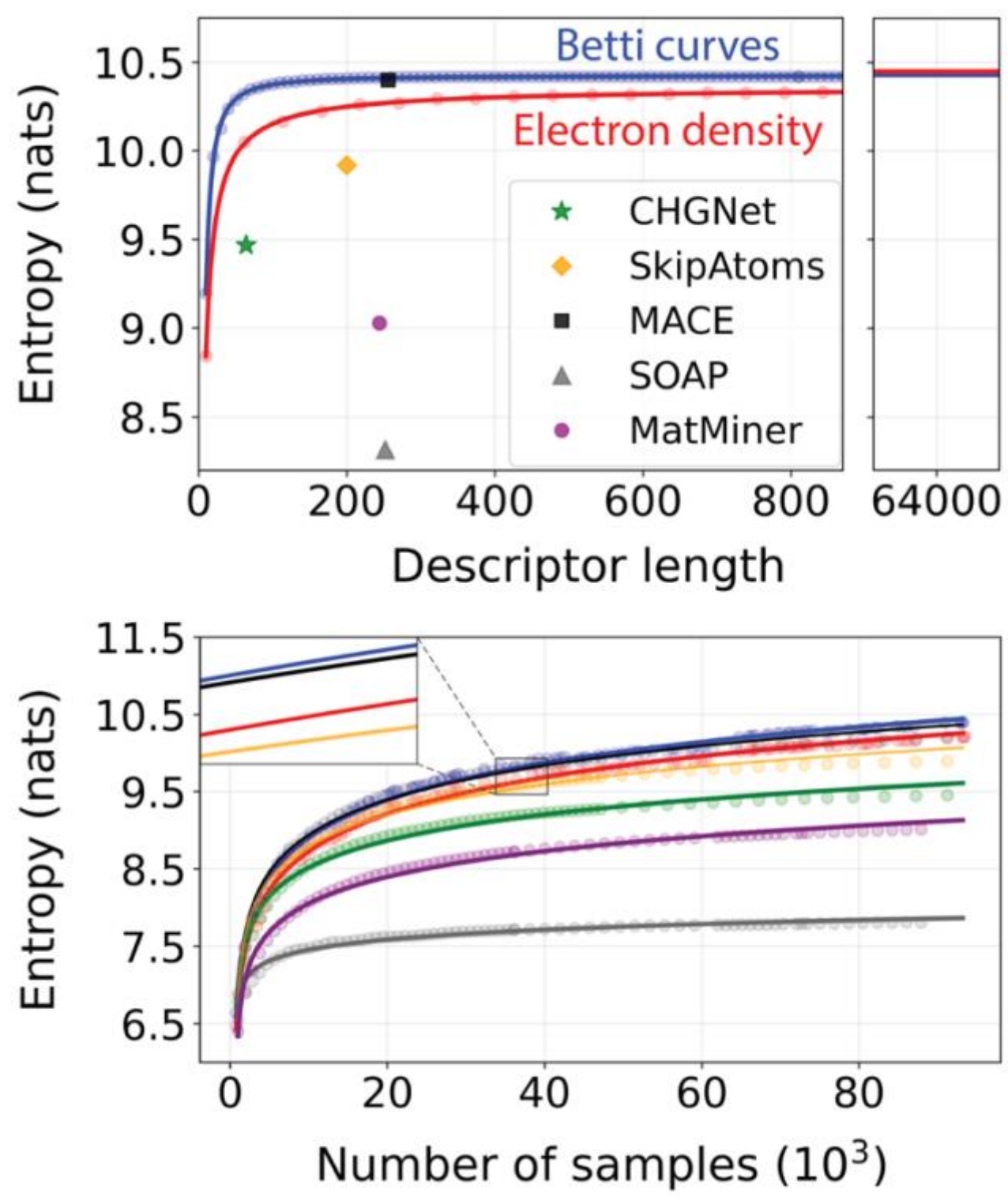

**Figure 5 | Shannon entropies of material representations.** (Top panel) Shannon entropy of electron densities and Betti curves generated from 93,178 materials with varied sampling density. The curves represent inverse power law fits of the computed data. Also shown are the entropies of five established descriptors with fixed length. (Bottom panel) Shannon entropies of these same descriptors computed on datasets with increasingly more samples. The curves represent logarithmic fits of these entropies.

For comparison, we computed the Shannon entropy of several established descriptors introduced in the previous section. Because these descriptors have fixed lengths and generally do not represent continuous features, resampling was not possible, so a single entropy value was computed for each. As shown by the markers in the top panel of **Figure 5**, feature embeddings from MACE achieve entropy values comparable to the maximum (~10.5 nats) obtained from electron densities and Betti curves. All other descriptors exhibit lower entropies, ranging from 8 nats (SOAP) to 10 nats (SkipAtom).

To further examine information content, we recomputed the entropy of each descriptor using datasets with increasing numbers of materials randomly drawn from the total of 93,178. Betti curves and electron densities were both standardized to descriptor lengths of 200, ensuring a consistent comparison and a comparable length with the previously established descriptors. As shown in the bottom panel of **Figure 5**, all descriptors exhibit a logarithmic increase in entropy as more samples are added. Two of the descriptors (Betti curves and MACE feature embeddings) converge to nearly identical entropy values across all dataset sizes, suggesting a potential upper limit to the information content of these materials. Nevertheless, how this information is represented clearly impacts ML performance, as highlighted by the varied classification accuracies in **Table 1**.

Our findings underscore the utility of topological data analysis (TDA), a powerful framework for extracting insights from complex datasets.[20,21] By focusing on shape and connectivity, TDA methods effectively capture subtle features (components, cycles, and voids) that often dictate properties of interest. These approaches are particularly effective when applied to high-dimensional data, offering interpretability that complements traditional ML techniques. In materials science, TDA has been used to predict formation[37] and defect[38] energies, quantify short-range order,[39] and characterize microstructure[40] by analyzing the persistent homology present in crystal structures and sample morphologies. Our work demonstrates that TDA can similarly be extended to the electron density, an intricate and high-dimensional quantity that is fundamental to all observable material properties. By processing this complex electron density into simpler topological representations through Betti curves, we enable ML models to more effectively harness this complexity.[41]

Despite the advantages shown in this work, applying TDA to electron density also comes with inherent limitations. Electron densities (and consequently Betti curves) lack phase information of the wavefunction, which is necessary to distinguish bonding from antibonding interactions.[42,43] As a result, while Betti curves may capture structural and bonding trends encoded in various electron densities, they cannot fully describe (anti)bonding interactions. A second limitation comes from the fact that electron densities are computationally demanding to compute using DFT. If such calculations have already been performed, many ground-state properties can often be extracted directly, reducing the need for ML. Nevertheless, Betti curves and related topological descriptors may offer broader applications beyond the prediction of ground-state properties that would otherwise be costly to obtain from DFT alone. The utility of these methods will also increase when paired with emerging ML approaches that can predict electron density at a fraction of the computational cost associated with typical DFT calculations.[44–47] Together, these advancements highlight the potential of combining topological descriptors with data-driven approaches to drive materials design.

## Code availability

The code used to query electron densities from the Materials Project, compute their Betti curves, and train classification models is provided at https://github.com/Bartel-Group/crystopo.

## Supporting Information

Additional details on the computational methods used in this work, including the generation of Betti curves from electron density, calculation of Bader charges and ICOBI values, structure fingerprint construction using Matminer, application of machine learning potentials, structure prototype classification, and Shannon entropy calculations. Includes supplementary figures comparing Betti curves across different cell sizes and correlating charge transfer with ICOBI values.

**Supplementary Information**

**Topological descriptors for the electron density of inorganic solids**

Nathan J. Szymanski[1], Alexander Smith[1], Prodromos Daoutidis[1], and Christopher J. Bartel[1,*]

**Table of contents**

[1] University of Minnesota, Department of Chemical Engineering and Materials Science, Minneapolis, MN, USA 55455

[*] Correspondence to cbartel@umn.edu

**Supplementary Note 1. Generation of Betti curves from electron density**

Betti curves were computed as topological descriptors of electron density using persistent homology, a mathematical framework that captures the connectivity of spatial structures. The following steps outline the methodology used to generate Betti curves from electron density grids obtained *via* density functional theory (DFT) calculations.

**1) Normalization and supercell generation**

Electron densities were extracted from the Materials Project[1] and normalized by unit cell volume to convert them into units of electrons per cubic angstrom (e/Å$^3$). These densities account for valence electron positions, with core electrons represented implicitly in the pseudopotential and excluded from the density. When core electrons are included (*e.g.*, when using pseudopotentials such as Li_sv or Mn_pv), they contribute to the density but generally have little effect on the Betti curves. As detailed below, these curves are calculated by focusing on changes to regions with low electron density ($\leq 2\ e/\text{Å}^3$) where bonding takes place, while core electrons are typically associated with higher densities. To account for the periodicity of the electron densities, and the effect that it has on their topology, supercells were created as follows:

- 3×3×3 supercells for primitive cells containing fewer than 10 atoms.
- 2×2×2 supercells for primitive cells containing 10–50 atoms.
- For structures with more than 50 atoms, the primitive unit cell was used directly.

In cases where a supercell was created, the electron density was tiled accordingly to match the enlarged structure. From preliminary testing, we found that further increases to the supercell size were unnecessary, having little effect on the resulting Betti curves. An example is provided in Supplementary Figure 2, which shows that changes to the Betti curves become negligible upon increasing cell size (beyond 8 atoms) for CdTe.

**2) Persistent homology and Betti number calculation**

To generate Betti curves from electron density, a series of superlevel sets built from regions of progressively lower electron density were added while tracking the formation and disappearance of topological features. These features were quantified using Betti numbers – describing connected components, cycles, and voids – computed for a sequence of filtration thresholds ($\rho_{\min}$). A set of 500 thresholds was defined within the range 0–2 e/Å$^3$, where each filtration step incrementally reduced the density threshold, adding to the superlevel set. To calculate the Betti numbers of each superlevel set, we used the CubicalComplex module in GUDHI.[2,3] The resulting Betti numbers were normalized by the total number of atoms in the supercell to account for size variations across different materials.

**3) Post-processing and Smoothing**

To enhance the interpretability of the Betti curves and reduce noise, a Gaussian smoothing function was applied with a standard deviation of $\sigma = 1$. The smoothing process was implemented using the gaussian_filter1d function in SciPy,[4] which convolves the Betti curves with a Gaussian kernel. Additionally, to prevent extreme fluctuations in the Betti numbers, an upper cap of 5 was enforced. Any portions of the Betti numbers exceeding this threshold were truncated.

## Supplementary Note 2. Calculation of Bader charges and ICOBI values

Density functional theory calculations based on the $r^2$SCAN meta-GGA functional[5] were performed using the Vienna Ab initio Simulation Package (VASP).[6,7] All structures were obtained from the Materials Project[1] and used as starting points for additional calculations. These relied on the use of projector-augmented wave potentials with an energy cutoff of 520 eV, and a Γ-centered k-point grid with spacing of 0.22 $Å^{-1}$. Projector-augmented wave (PAW) pseudopotentials were chosen based on recommendations from the Materials Project. Structure relaxations were carried out using convergence criteria of $10^{-6}$ eV for electronic optimizations and 0.01 eV/Å for ionic optimizations. Following relaxation, static calculations were performed using twice the default number of electronic bands (NBANDS) to ensure sufficient coverage of empty states. The LAECHG, LCHARG, and LASPH flags were all set to True during these calculations, generating the necessary charge density files for post-processing. Spin polarization was included for materials containing any magnetic elements. These moments were initialized in a ferromagnetic configuration for all compounds.

Bader charge analysis[8] was used to describe ionic bonding, as it quantifies the transfer of electrons between atoms. This method partitions the electron density into distinct atomic basins, assigning each region to the nearest local maximum in the electronic density. By integrating the number of electrons present within each basin, Bader charge analysis provides a measure of the electronic charge associated with each atom.

The Integrated Crystal Orbital Bond Index (ICOBI)[9] was computed using LOBSTER[10] to assess covalent bonding interactions. Unlike Crystal Orbital Hamilton Population (COHP), which provides an energy-resolved measure of bonding and antibonding interactions,[11] ICOBI is derived from the direct overlap of projected atomic orbitals and represents a measure of bond order in periodic systems. Higher ICOBI values indicate stronger covalent bonding, while lower values suggest more ionic bonding. From the DFT calculations described above, ICOBI analysis was performed using the pbeVaspFit2015 basis set, considering energies ranging from -20.0 eV to 5.0 eV with 2000 points sampled and a Gaussian smearing width of 0.2 eV. The orbitals included in each basis set were chosen based on the POTCAR files used in our VASP calculations – following the predefined pbeVaspFit2015 standard basis set – as implemented in the pymatgen.io.lobster module.[12,13] ICOBI values were extracted directly from the overlap matrix generated by these LOBSTER calculations and used to assess ICOBI in the rocksalt materials discussed throughout the main text.

**Supplementary Note 3. Structure fingerprint construction using Matminer**

To encode structural information into vectors suitable for machine learning models, we used the *SiteStatsFingerprint* and *CrystalNNFingerprint* classes from matminer.[14] The fingerprinting strategy first characterizes each site in a structure according to its local coordination environment. Specifically, *CrystalNNFingerprint* identifies the nearest neighbors of a site by combining Voronoi-based geometry with heuristics for chemical bonding to probabilistically assign coordination numbers and compute local order parameters. These order parameters include a coordination likelihood and structural motifs such as tetrahedral or octahedral environments, yielding a set of features that describe how closely the local neighborhood of each site resembles idealized geometries. The *SiteStatsFingerprint* method transforms these site-level features into structure-level descriptors by computing statistical aggregates (the mean, standard deviation, minimum, and maximum) across all atomic sites in a structure for each coordination feature. This yields a fixed-length numerical vector that represents the distribution of local environments in the structure. The structure-level descriptors are invariant to permutation of atomic indices and system size, allowing direct comparison between different structures and robust input to models such as the random forests used in our study.

**Supplementary Note 4. Pre-trained machine learning potentials**

We used feature embeddings resulting from two pretrained machine-learned interatomic potentials (MLIPs) developed in prior work. The first (CHGNet)[15] is a charge-aware graph neural network (GNN) trained to predict energy, forces, stress, and magnetic moments for inorganic solids. The model takes as input only the atomic identities and positions and infers latent charge information from magnetic moments during training. CHGNet builds atom graphs and bond graphs to incorporate both pairwise and angular information through message passing. It explicitly regularizes latent features with magnetic moments to capture charge-dependent chemical behavior. The model was trained on the Materials Project Trajectory (MPtrj) dataset, which contains ~1.5 million DFT-computed configurations from structure relaxations and static calculations. Training was performed using an 8:1:1 train-validation-test split and achieved mean absolute errors of about 30 meV/atom (energy), 77 meV/Å (forces), 0.348 GPa (stress), and 0.032 $\mu_B$ (magnetic moment) on the test set. To extract feature embeddings from this pretrained model (available at https://github.com/CederGroupHub/chgnet), we set *return_crystal_feas* to True when performing a forward pass (using *model._compute*). This takes the embeddings obtained from convolution and pools them across all atoms in the structure to produce a single crystal-level feature vector, which we then used as input to the random forests described in the main text.

The second model, MACE-MP-0a medium,[16] is based on the MACE architecture[17] – an equivariant graph tensor network derived from the atomic cluster expansion framework. This model uses two message-passing layers with 4-body equivariant features and up to angular momentum $l_{max} = 3$, trained with 128 channels per tensor decomposition and a radial cutoff of 6 Å. MACE-MP-0 was trained on the same MPtrj dataset as CHGNet, with ~1.5 million configurations. Unlike CHGNet, MACE does not incorporate charge or magnetic information explicitly but provides architectural innovations like the use of high-body-order equivariant features. The model used in this work (MACE-MP-0) achieved mean absolute errors of 20 meV/atom (energy) and 45 meV/Å (force). To extract feature embeddings from this pretrained model (available at https://github.com/ACEsuit/mace), we used the *get_descriptors* method in the *MACECalculator* class with *invariants_only* set to True and *num_layers* set to $-1$. With these settings, a forward pass is made through the model and node-level features are extracted from all interaction layers, yielding per-atom descriptors that capture local atomic environments while preserving physical symmetries (being invariant to rotation). These descriptors were then flattened and pooled across all atoms in the structure to produce a single crystal-level feature vector, which we then used as input to the random forests described in the main text.

**Supplementary Note 5. Classification of structure prototypes**

To map structures from the Materials Project database to crystallographic prototypes from the AFLOW Encyclopedia of Crystallographic Prototypes,[18–21] we employed the *StructureMatcher* class from pymatgen.[12] This automated approach allows for the efficient processing of a large number of structures and identification of their corresponding prototypes in a high-throughput manner. The structure matching process was implemented as follows:

1. For each prototype in AFLOW, we selected a reference structure.
2. Each reference structure was compared to candidate structures from the Materials Project using the *StructureMatcher* framework.
3. To focus solely on structural similarities while ignoring compositional differences, we utilized the *FrameworkComparator* class within *StructureMatcher*.
4. All tolerance parameters were set to their default values:
    a. Length tolerance (*ltol*) = 0.2, expressed as a fractional value that allows for a 20% deviation in lattice parameters.
    b. Site tolerance (*stol*) = 0.3, defined as the fraction of the average free length per atom $(V/N)^{1/3}$, permitting a 30% deviation in normalized site distances.
    c. Angle tolerance (*atol*) = 5°, allowing up to 5° deviation in lattice angles.

For each prototype, we considered only matches that had electron density data available, resulting in a dataset of 93,178 structures mapped to 846 crystallographic prototypes. While these mappings can sometimes be sensitive to the chosen tolerance parameters, particularly for borderline cases, we maintained a one-to-one mapping of materials to prototypes to ensure a well-defined machine learning classification task.

**Supplementary Note 6. Calculation of Shannon entropy**

The Shannon entropy ($H$) for $n$ samples, encoded using descriptors of the form $X_i$, is given by the equation:

$$H(\{X\}) = -\frac{1}{n}\sum_{i=1}^{n} \log\left[\frac{1}{n}\sum_{j=1}^{n} K_h(X_i, X_j)\right] \quad (1)$$

Where $K_h$ is a Gaussian kernel defined as:

$$K_h(X_i, X_j) = \exp\left(\frac{-\|X_i - X_j\|^2}{2h^2}\right) \quad (2)$$

Here, we estimate the bandwidth ($h$) using Silverman's rule of thumb, based on the median absolute deviation of the descriptors. Specifically, we calculate the initial bandwidth as:

$$h = \left(\frac{4}{d+2}\right)^{1/(d+4)} \times \sigma n^{-1/(d+4)}$$

Where $d$ is the dimensionality of the data, $n$ is the sample size, and $\sigma$ is the mean standard deviation across dimensions. This bandwidth is then refined using leave-one-out cross-validation for optimal density estimation.

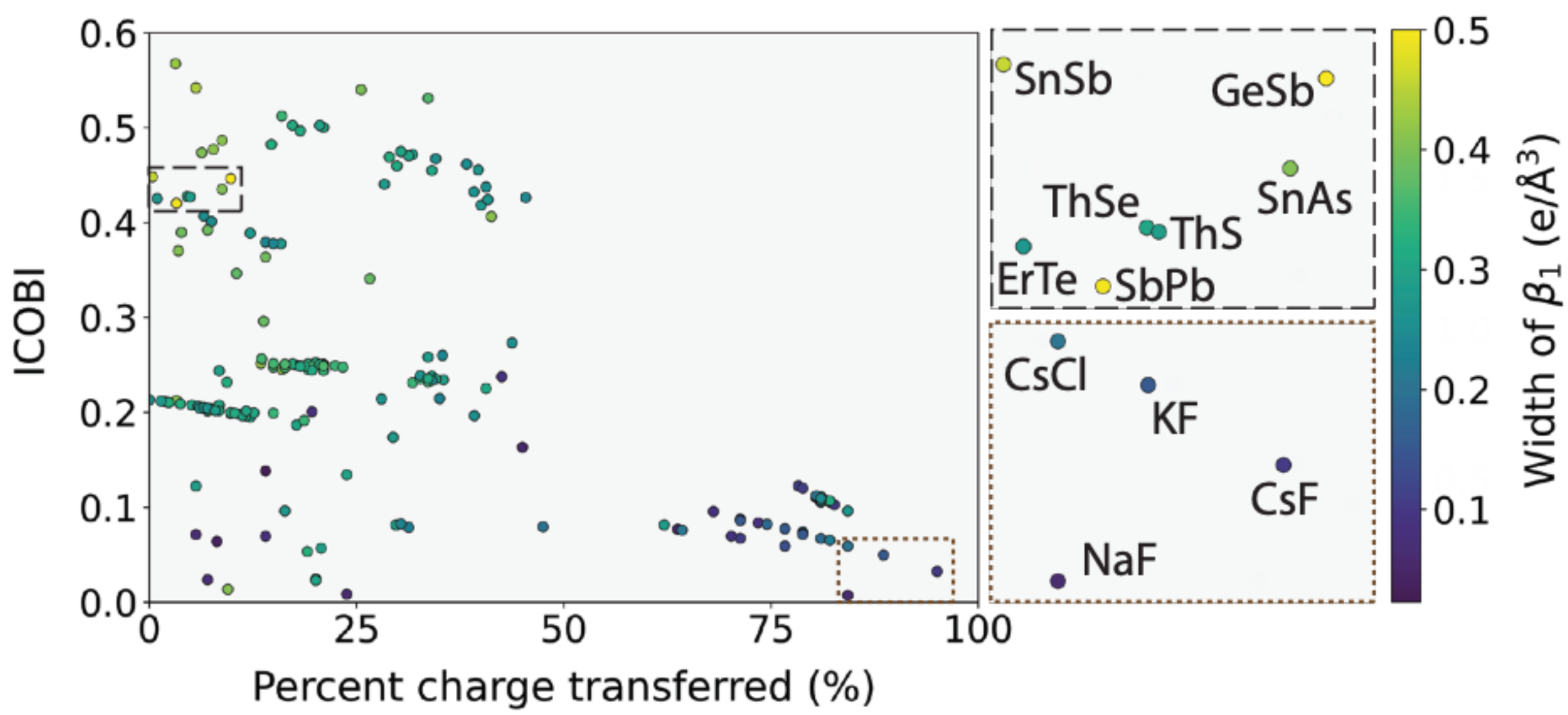

**Supplementary Figure 1.** Scatter plot showing the relationship between the percent of charge transferred (x-axis) and ICOBI (y-axis) for various rocksalt materials. Each point represents a distinct material, with its color indicating the width of $\beta_1$ in its Betti curves. In contrast to Figure

3 in the main text, the values for percent charge transferred are derived from Mulliken, rather than Bader, charges.

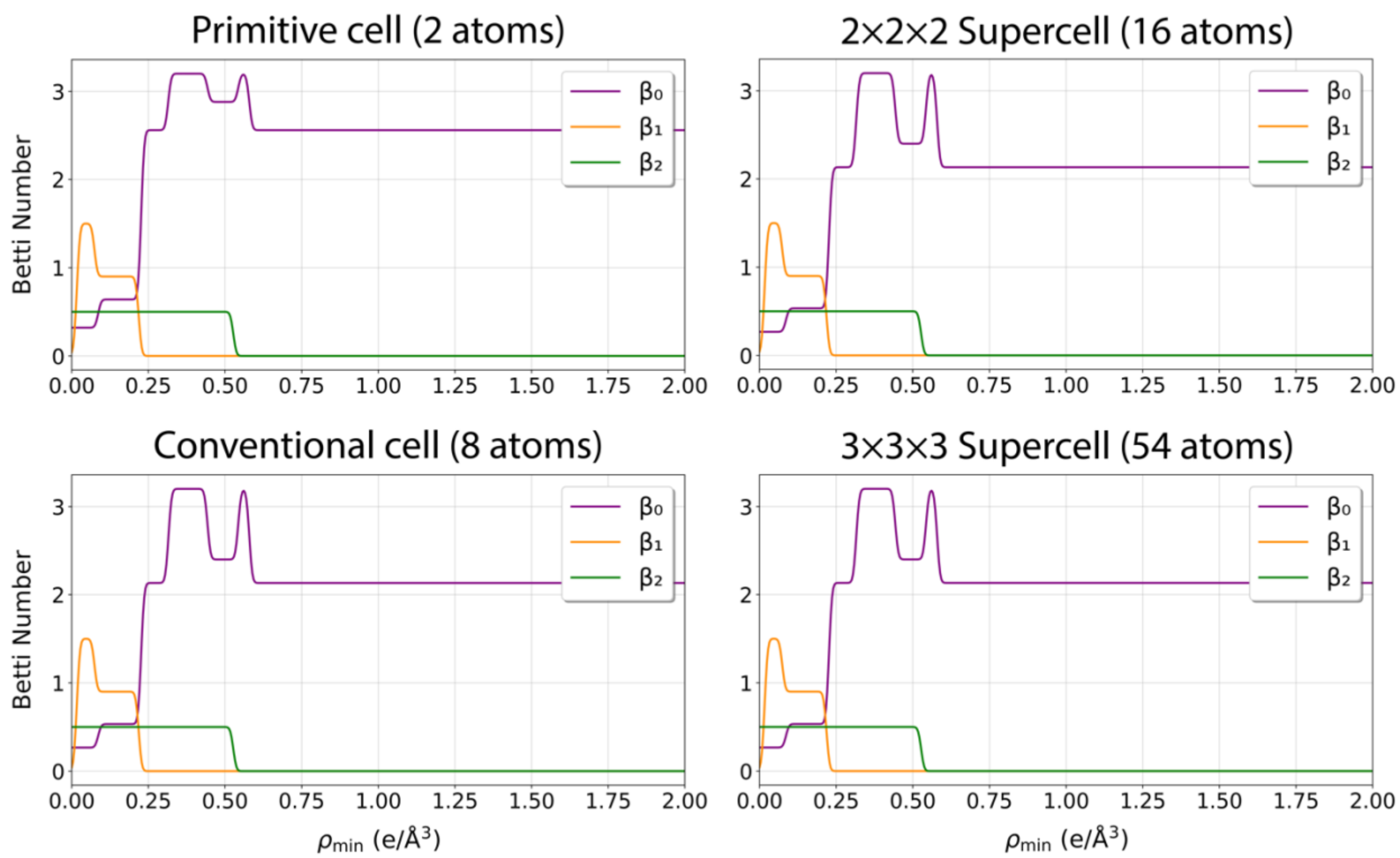

**Supplementary Figure 2.** Betti curves derived from CdTe using varied cell representations. For the purpose of comparison in these examples, we did not tile the electron density and instead just directly assess the electron density of the calculated cells.